\documentstyle[12pt]{article}
\newcommand{\bec}{\begin{center}}
\newcommand{\ec}{\end{center}}
\newcommand{\bee}{\begin{equation}}
\newcommand{\ee}{\end{equation}}
\newfont{\blackboard}{msbm10 scaled\magstep2}
\newcommand{\Z}{\mbox{\blackboard\symbol{"5A}}}

\begin{document}
\large
\begin{titlepage}
\bec
{\Large\bf  D-Branes, AdS/CFT Correspondence,\\
and T-Duality  \\}
\vspace*{15mm}
{\bf Yuri Malyuta \\}
\vspace*{10mm}
{\it Institute for Nuclear Research\\
National Academy of
Sciences of Ukraine\\
252022 Kiev, Ukraine\\}
e-mail: interdep@kinr.kiev.ua\\
\vspace*{35mm}
{\bf Abstract\\}
\ec
The $D$-brane spectra
in the Type IIB string theory
compactified on $AdS_{p+2}\times S^{8-p}$
are computed using the K-theory
approach. The  results signal 
the existence of the 
mirror-symmetry-analogue
for $D$-branes, analogous 
to that rised in the context 
of derived categories. 
\vspace*{1cm}
\end{titlepage}
\section{Introduction}
$D$-branes play a significant role in 
superstrings and superconformal
field theories.
Two of the most outstanding 
developments in this direction have been
achieved:\\
1. The generalized AdS/CFT correspondence
\cite{1.}, which relates the 
superconformal field theory on
$Dp$-branes placed at the orbifold 
singularity and the Type IIB string
theory compactified on 
$AdS_{p+2}\times H^{8-p}$\ ;\\
2. The K-theory approach to $D$-brane
charges \cite{2.,3.,4.}, which identifies 
$D$-brane charges with elements
of Grothendieck K-groups \cite{5.,6.,7.}
of horizon manifolds.

	In the present paper we use
K-theory to compute the 
$D$-brane spectra in the Type IIB
string theory compactified on 
$AdS_{p+2}\times S^{8-p}$. 

\section{$D$-brane spectra}
Let us consider the fibre bundle
\vspace*{5mm}
\bec
\begin{tabular}{ccc}
$S^{8-p}$\hspace*{-0.2cm}&$\rightarrow$ &
\hspace*{-1.3cm}$B^{9-p}$ \\
&    &\hspace*{-2cm}$\downarrow$ \\
 &   &\hspace*{-2mm}$B^{9-p}/S^{8-p}$  \\
\end{tabular}
\ec
The K-groups characterizing this 
bundle are related by the exact 
hexagon \cite{7.}
\vspace*{5mm}
\bec
\begin{tabular}{ccccccc}
&&\hspace*{-1cm}$\widetilde{K}(B^{9-p}/S^{8-p})$
&\hspace*{-4cm}$\rightarrow$
&\hspace*{-6.2cm}$\widetilde{K}(B^{9-p})$&&\\
\hspace*{2.7cm}$\stackrel{\delta}{\nearrow}$
&& &\hspace*{5cm} & &&\hspace*{-5cm}$\searrow$\\
$\widetilde{K}(SS^{8-p})$& & & & & 
& \hspace*{-2.5cm}$\widetilde{K}(S^{8-p})$\\
\hspace*{2.7cm}${\nwarrow}$&& & & &
&\hspace*{-5cm}$\stackrel{\delta}{\swarrow}$\\
&&\hspace*{-2cm}$\widetilde{K}(SB^{9-p})$
&\hspace*{-6.3cm}$\leftarrow$
&\hspace*{-7cm}$\widetilde{K}(S(B^{9-p}/S^{8-p}))$&&\\
\end{tabular}
\ec
\vspace*{5mm}
where $\delta$ is the coboundary homomorphism.
This hexagon is the counterpart of the
generalized AdS/CFT correspondence (cf. \cite{3.}).

	Since

\[\widetilde{K}(B^{9-p})=\widetilde{K}(SB^{9-p})=0 \ ,\]

\hspace*{-6mm}the hexagon splits into the exact sequences 
\bee
0\rightarrow\widetilde{K}(SS^{8-p})\stackrel{\delta}
{\rightarrow}
\widetilde{K}(B^{9-p}/S^{8-p})\rightarrow 0
\ee
\bee
0\rightarrow\widetilde{K}(S^{8-p})\stackrel{\delta}
{\rightarrow}
\widetilde{K}(S(B^{9-p}/S^{8-p}))\rightarrow 0
\ee

\hspace*{-6mm}The sequences (1) and (2) are 
related by T-duality.

	The group $\widetilde{K}(SS^{8-p})$ from (1)
reproduces the $D$-brane spectrum
\vspace*{7mm}
\bec
Table 1\\
\vspace*{7mm}
\begin{tabular}{|c|c|c|c|c|c|c|c|c|c|c|c|}
\hline
\normalsize$Dp$ &\normalsize$D9$
&\normalsize$D8$ &\normalsize$D7$
&\normalsize$D6$ &\normalsize$D5$
&\normalsize$D4$ &\normalsize$D3$
&\normalsize$D2$ &\normalsize$D1$
&\normalsize$D0$ &\normalsize$D(-1)$
\\  \hline
\normalsize$S^{9-p}$&\normalsize$S^{0}$
&\normalsize$S^{1}$ &\normalsize$S^{2}$
&\normalsize$S^{3}$ &\normalsize$S^{4}$
&\normalsize$S^{5}$ &\normalsize$S^{6}$
&\normalsize$S^{7}$ &\normalsize$S^{8}$
&\normalsize$S^{9}$ &\normalsize$S^{10}$
\\ \hline
\normalsize$\widetilde{K}(S^{9-p})$
&\normalsize$\Z$ &\normalsize0
&\normalsize$\Z$ &\normalsize0
&\normalsize$\Z$ &\normalsize0
&\normalsize$\Z$ &\normalsize0
&\normalsize$\Z$ &\normalsize0
&\normalsize$\Z$  \\ \hline
\end{tabular}\\
\ec
\vspace*{5mm}
which coincides with the known result in Type IIB 
theory \cite{8.}.

	        The group $\widetilde{K}(S^{8-p})$ 
from (2)
reproduces the $D$-brane spectrum
\bec
Table 2\\
\vspace*{7mm}
\begin{tabular}{|c|c|c|c|c|c|c|c|c|c|c|c|}
\hline
\normalsize$Dp$ &\normalsize$D9$
&\normalsize$D8$ &\normalsize$D7$
&\normalsize$D6$ &\normalsize$D5$
&\normalsize$D4$ &\normalsize$D3$
&\normalsize$D2$ &\normalsize$D1$
&\normalsize$D0$ &\normalsize$D(-1)$
\\  \hline
\normalsize$S^{8-p}$&\normalsize$S^{-1}$
&\normalsize$S^{0}$ &\normalsize$S^{1}$
&\normalsize$S^{2}$ &\normalsize$S^{3}$
&\normalsize$S^{4}$ &\normalsize$S^{5}$
&\normalsize$S^{6}$ &\normalsize$S^{7}$
&\normalsize$S^{8}$ &\normalsize$S^{9}$
\\ \hline
\normalsize$\widetilde{K}(S^{8-p})$
&\normalsize0 &\normalsize$\Z$ 
&\normalsize0 &\normalsize$\Z$ 
&\normalsize0 &\normalsize$\Z$ 
&\normalsize0 &\normalsize$\Z$ 
&\normalsize0 &\normalsize$\Z$ 
&\normalsize0 \\ \hline
\end{tabular}\\
\ec
\vspace*{5mm}
which signals the existence 
of the mirror-symmetry-analogue
for branes, analogous to that rised in the
context of derived categories \cite{4.}. 
\section{Vacuum manifold}
Using standard definitions
\cite{7.}, we obtain
\[\widetilde{K}(SS^{8-p})=\pi_{9-p}(BU) \ ,\]
\[\widetilde{K}(S^{8-p})=\pi_{8-p}(BU) \ ,\]
where $BU$ is the inductive limit
of the manifold
\bee
U(2N)/U(N)\times U(N)
\ee

        The vacuum manifold (3) has the following
interpretation in terms of $D$-branes \cite{9.}.
When $2N$ coinciding branes are separated
to form two parallel stacks of $N$ coinciding
branes, their gauge symmetry $U(2N)$ is
spontaneously broken to $U(N)\times U(N)$.
This situation generically allows for the
existence of topological solitons.

\section {Acknowledgements}

\hspace*{6mm}I would like to thank S. Gukov for
stimulating discussions. It is a pleasure
to thank E. Witten for his attention to
this work.

\end{document}